\documentstyle[11pt]{article}

\title{\bf Triple pomeron vertex in the limit $N_c\rightarrow\infty$.}
\author{M.Braun$^{a,b}$ and G.P.Vacca$^a$ \\
$^a$Department of Physics,
University of Bologna and INFN, Italy, and\\
$^b$Department of High Energy Physics, University of S.Petersburg }
 \date{October 1997}
\def\beq{\begin{equation}}
\def\eeq{\end{equation}}
\def\noi{\noindent}
\begin{document}
\maketitle
\medskip
\noi{\bf Abstract}

In the hard pomeron theory with the number of colours $N_c\rightarrow\infty$
the diffractive amplitude obtained in [3] is compared with the results
found for $N_c=3$ in [1] and in the dipole approach in [5]. It is shown
that the double pomeron exchange contribution can be substituted by
an equivalent triple pomeron interaction term. After such a substitution
the triple pomeron vertices in [1,3,5] essentially coincide. It is
demonstrated that, in any form, the triple pomeron vertex is conformal
invariant. It is also shown that higher order densities in the
dipole approach do not involve 1 to k pomeron verteces with $k>2$ but
are rather given by a set of pomeron fan diagrams with only a triple
pomeron coupling.

\newpage
\section{Introduction}
In the hard pomeron theory the first step towards unitarization goes
through the construction of the amplitude generated by the exchange
of four reggeized gluons. In the old Regge-Gribov theory this amplitude
was a sum of the double pomeron exchange (DPE) and 
triple pomeron interaction (TPI) contributions, both having essentially the same
behaviour at high energies. For a realistic case of the number of colours
$N_c=3$ the four-gluon system was studied by J.Bartels and M.Wuesthoff [1].
They obtained a complicated system of coupled equations for different
colour channels amplitudes. The inhomogeneous terms for this system
included a structure which was interpreted as a triple-pomeron vertex.
Its explicit expression, found in [1] is rather complicated (it consists
of 19 terms). Later it was demonstrated that the found vertex was
conformal invariant [2]. In a recent publication [3]  we repeated the
 derivation of [1] in the limit $N_c\rightarrow\infty$ guided by the idea that
 in this limit the leading contribution reduces to a single BFKL pomeron [4].
 We found that in the limit $N_c\rightarrow\infty$ the complicated system
 of J.Bartels and M.Wuesthoff decouples and can be explicitly solved both
 in the leading and subleading approximations in $1/N_c$. The leading
 contribution  indeed reduces to a single BFKL pomeron exchange, as expected.
 The subleading diffractive amplitude was found to be a sum of two terms,
 the DPE and TPI, in full correspondence with the Regge-Gribov picture.
 However the found triple pomeron vertex (also quite complicated) resulted
 different from the one obtained by J.Bartels and M.Wuesthoff. Also its
 conformal properties remained unclear. On the other hand, at the same time
 R.Peschanski calculated the double dipole density in the A.Mueller dipole
 approach [5], valid in the $N_c\rightarrow\infty$ limit. From his result he
 extracted the triple pomeron vertex, which is rather simple and superficially
 different from the ones discussed above, obtained via the $s$-channel
 unitarity approach. No contribution which could be interpreted as the DPE
 seems to appear in the dipole approach.

 Given the variety of the expressions for the 4-gluon diffractive
 amplitude and the triple pomeron vertex, we dedicate this note to compare
 these different results in the $N_c\rightarrow\infty$ limit. Our main
 conclusion is that, in fact, they essentially coincide, since, as will be
 explicitly demonstrated, the DPE contribution can be substituted by
 a completely equivalent TPI term (but not {\it vice versa}). Once this is
 done, our vertex found in [3] coincides with the one in [1], provided one
 takes the limit $N_c\rightarrow\infty$. Coupled to pomerons, this vertex
 effectively reduces to the one found by R.Peschanski in [5]. However this
 does not mean that the double dipole density in the dipole approach
 coincides with the diffractive amplitude in the $s$-channel unitarity
 approach: there are certain terms in the latter which are missing in the
 double dipole density.

 As a byproduct of our study we prove that the triple pomeron vertex
 found in [3] is also conformal invariant. We also comment on the higher-
 order dipole densities in A.Mueller's approach, in relation to the
 form of $1\rightarrow k$ pomeron vertex proposed by R.Peschanski in [5].

 The contents of this note is distributed as follows. In Sec. 2, of an
 introductory character, we  present a generalization of the 4-gluon
 amplitude in the $N_c\rightarrow\infty$ limit to a non-forward direction,
 necessary to study its conformal properties. Sec. 3 is devoted to these
 properties. In Sec. 4 we demonstrate the equivalence of the DPE and
 certain TPI terms. In Sec.5 we compare the 4-gluon diffractive amplitudes
 found in different approaches. In Sec. 6 we briefly discuss the higher
 order dipole densities. Finally Sec. 7 contains some conclusions.

 \section{Four reggeized gluons with a nonzero total momentum}

 Since the conformal transformations do not conserve the total momentum
 of the gluons, to study conformal properties of the amplitudes generated
 by the exchange of 4 gluons in the limit $N_c\rightarrow\infty$ we have
 to generalize the derivation presented in [3] to the case when the gluons have
 their total momentum different from zero. This generalization is quite
 straightforward and the main change will concern the notations, which in
 [3] essentially used the fact that the total momentum is zero. The colour
 structure and the derivation lines remain the same, so that we shall be quite
 brief, just presenting the results.

 The basic quantity is the amplitude $D_2$ corresponding to the exchange of
 two reggeized gluons (the BFKL amplitude). It satisfies the BFKL equation
\beq
S_{20}D_{2}=D_{20}+g^{2}N_cV_{12}D_{2}
\eeq
where 
$ S_{20} $ is the 2 gluon  "free"  Schroedinger operator for the energy 
$ 1-j $
\beq
S_{20}=j-1-\omega(1)-\omega(2)
\eeq
$ \omega(k) $ is the gluon Regge trajectory  and 
$ V_{12} $ is the BFKL interaction. We use the notation in which only the
number of the gluon is indicated whose momentum enters as a variable. The
inhomogeneous term for the non-forward direction and $N_c\rightarrow\infty$ is
\beq
D_{20}(1,2)=D_{20}(2,1)=g^2N_c\left(f(1+2,0)-f(1,2)\right)
\eeq
where $f(1,2)=f(2,1)$ is a contribution of the $q\bar q$ loop with gluon 1 attached
to $q$ and gluon 2 attached to $\bar q$. Its explicit form can be easily found
(see Appendix) but
has no importance for the following.

The 3 gluon amplitude $D_3$, as in the forward case, is found to be
constructed in terms of $D_2$:
               \beq 
D_{3}^{(123)}=-D_{3}^{(213)}=g\sqrt{N_c/8}(D_{2}(2,1+3)-D_{2}(1,2+3)
-D_{2}(3,1+2))
\eeq
where the upper indeces 123 and 213 show the order of the gluons along the
 $q\bar q$ loop.

The 4-gluon amplitudes $D_4$ in the leading approximation in $1/N_c$
correspond to neighbour gluons being in the adjoint colour state and all
the gluons lying on the surface of a cylinder attached to the $q\bar q$ loop.
There are two independent amplitudes of this type, corresponding to the
order of the gluons 1234 and 2134. Both are found to be expressed via the
BFKL amplitudes $D_2$, so that the contribution of the 4-gluon exchange
reduces to a single BFKL pomeron. Explicitly one obtains in the same
manner as in [3]
\beq
D_{4}^{(1234)}=(1/4)g^{2}N_c(D_{2}(1,2+3+4)+D_{2}(4,1+2+3)-D_{2}(1+4,2+3))
\eeq 
and
\beq
D_{4}^{(2134)}=(1/4)g^{2}N_c(D_{2}(2,1+3+4)+D_{2}(3,1+2+4)-
D_{2}(1+2,3+4)-
D_{2}(1+3,2+4))
\eeq

The diffractive amplitude $D_4^{(0)}$, which is of main interest to us,
corresponds to pairs of gluons 12 and 34 being in the vacuum colour state.
It is subleading in $1/N_c$ and satisfies an equation
\beq
S_{40}D_{4}^{(0)}=D_{40}^{(0)}+D_{2\rightarrow 4}^{(0)}+D_{3\rightarrow 
4}^{(0)}+D_{4\rightarrow 4}^{(0)}+g^{2}N_c(V_{12}+V_{34})D_{4}^{(0)}
\eeq
Here $S_{40}=j-1-\sum_{i=1}^4\omega(i)$,
\beq
D_{40}^{(0)}=\frac{1}{2}g^2(\sum_{i=1}^4 D_{20}(i,1+2+3+4-i)-
\sum_{i=2}^4D_{20}(1+i,2+3+4-i))
\eeq
Terms $D_{2\rightarrow 4}^{(0)}$,... etc. come from transitions into the
4-gluon diffractive state from states with 2,...etc gluons in the leading
(cylinder) configuration. As in [1,3] their sum can be presented as a
certain operator $Z$ (the three-pomeron vertex) acting on the BFKL
pomeron:
\beq
D_{2\rightarrow 4}^{(0)}+D_{3\rightarrow 
4}^{(0)}+D_{4\rightarrow 4}^{(0)}=ZD_2
\eeq

>From the explicit form of the contributions on the left-hand side,
which can be found using the $s$-channel unitarity relations, one finds
   \[  
Z\,D_{2}=(1/2)g^{2}\]\[
[2G(1,3+4,2)+2G(3,1+2,4)+G(1,2+4,3)+G(1,2+3,4)+G(2,1+4,3)+G(2,1+3,4)\]\[
-G(1,4,2+3)-G(1,3,2+4)-G(2,4,1+3)-G(2,3,1+4)-G(3,2,1+4)-
G(3,1,2+4)\]\beq
-G(4,2,1+3)-G(4,1,2+3)+
G(2+3,0,1+4)+G(1+3,0,2+4)]
\eeq
In this expression function $G(1,2,3)$ is defined as the vertex
$K_{2\rightarrow 3}$ for the transition of 2 to 3 gluons, integrated with
the BFKL pomeron and regularized in the infrared by terms proportional to
the gluon trajectory in the same manner as in the total BFKL kernel:
\[
G(1,2,3)=G(3,2,1)=\]\beq-g^2 N_c\, W(1,2,3)-
D(1,2+3)(\omega(2)-\omega(2+3))-
D(1+2,3)(\omega(2)-\omega(1+2))
\eeq
where
\beq
W(1,2,3)=\int \frac{d^2k'_1}{(2\pi)^3}
K_{2\rightarrow 3}(1,2,3;1',3')
D(1',3')
\eeq
and the kernel $K_{2\rightarrow 3}$ is given by
\[
K_{2\rightarrow 3}(k_1,k_2,k_3;q_1,q_3)=\]\beq
-\frac{(k_2+k_3)^2}{(q_1-k_1)^2q_3^2}
-\frac{(k_1+k_2)^2}{q_1^2(q_3-k_3)^2}+\frac{k_2^2}{(q_1-k_1)^2
(q_3-k_3)^2}+\frac{(k_1+k_2+k_3)^2}{q_1^2q_3^2}
\eeq

Eq. (7) can be easily solved. Evidently the solution may be constructed as a
sum of two terms corresponding to the two parts of the inhomogeneous
term $D_{40}^{(0)}$ and $ZD_2$:
\beq
D_4^{(0)}=D_4^{DPE}+D_4^{TPI}
\eeq
The term $D_4^{DPE}$ coming from the inhomogeneous term $D_{40}^{(0)}$ is the
DPE contribution. Its explicit form can be conveniently written using the
quark loop density in the transverse coordinate space defined by the Fourier
transform (see Appendix)
\beq
f(1,2)=\int d^{2}r \rho_l(r)e^{ik_1r}
\eeq
where $l=k_1+k_2$ . Then one finds
\beq
D_{4}^{DPE}=(1/4)g^{4}N_c\int d^{2}r \rho_l(r)D_{4}^{(r)}
\eeq
Here
$D_{4}^{(r)}$ is a convolution in the "energy" 
$ 1-j $ of two independent BFKL pomerons
\beq
D_{4}^{(r)}=\int dj_{12} 
dj_{34}\delta(j-j_{12}-j_{34})D_{2,j_{12}}^{(r)}(1,2) 
D_{2,j_{34}}^{(r)}(3,4) \eeq where the pomeron $ D_{2,j}^{(r)}(1,2) $
satisfies the equation
\beq
S_{20}D_{2,j}^{(r)}=\prod_{j=1}^{2}(e^{ik_{j}r}-1) 
+g^{2}N_cV_{12}D_{2,j}^{(r)}
\eeq
and similarly for the second pomeron.

The part $D_4^{TPI}$ is the TPI contribution.  It can be written
 as a convolution   in the rapitidy space:
 \[
 D_4^{TPI}(1,2,3,4;Y)=
 \]\beq\int_{0}^YG_2(1,2;1'2';Y-y)G_2(3,4,;3'4';Y-y)
 \otimes Z(1',2',3',4';1'',2'')\otimes D_2(1'',2'';y)
 \eeq
 where $G_2$ is the BFKL Green function and the symbols $\otimes$ mean
 integrations over intermediate momenta. This equation clearly shows that
 $Z$ is just the three-pomeron vertex. Its explicit form can be read from
 Eq. (10). As compared to the forward case studied in [3],
  the only difference is the appearance of a new independent argument in
  functions $G$.

  \section{Conformal invariance}

  In [2] it was shown that a vertex $V(1234)$ defined by a relation
  similar to (10)
  \[
  V(1234)D_2=\frac{1}{2}g^2(G(1,2+3,4)+G(2,1+3,4)+G(1,2+4,3)+G(2,1+4,3)
  \]\beq-
  G(1+2,3,4)-G(1+2,4,3)-G(1,2,3+4)-G(2,1,3+4)+G(1+2,0,3+4))
  \eeq
  is conformal invariant in the following sense. If one transforms
 $ VD_2 $ to the transverse coordinate space and integrates it
 over the 4 gluon coordinates with a
 conformal invariant function, the resulting integral is invariant under
 conformal transformation of gluon coordinates. Comparing (20) and (10)
  we observe that our vertex $Z$ is just a sum of permutations of gluons
 in $V$
  \beq
  Z=V(1324)+V(1423)
  \eeq
 Then the conformal invariance of $Z$ trivially follows from the
 conformal invariance of $V$, proven in [2].

 In the rest part of this section we are going to demonstrate a stronger
 result: not only the combination (20) of functions $G$ is conformal invariant,
 but each function $G(1,2,3)$ is conformal invariant by itself. So this
 function represents a natural generalization of the BFKL kernel not only
 in respect to its infrared stability but also in its conformal properties.

 The proof of the conformal invariance of $G(1,2,3)$ is straightforward. The
 main technical problem is its transformation to the coordinate space. This
 task was actually already solved in [2], although there the transformation was
 applied to the vertex $V$ as a whole, integrated over the coordinates, which
 resulted in certain complications. Having this in mind, we shall present
 only the final expressions for $G(1,2,3)$ in the coordinate space 
 with some comments.

 Denote the integral part of $G$ in (11) as $G_1$ and the rest terms with
 the gluon trajectories as $G_2$. Transformation of $G_1$ to the coordinate
 space is straightforward and gives
 \beq
 G_1(r_1,r_2,r_3)=A_1D_2(r_1,r_3)
 \eeq
 where $A_1$ is an operator in the coordinate space
 \[
 A_1=\frac{g^2N_c}{8\pi^3}[
  2\pi\delta^2(r_{23})\nabla_3^2(c-\ln r_{13})\nabla_3^{-2}+
  2\pi\delta^2(r_{12})\nabla_1^2(c-\ln r_{13})\nabla_1^{-2}\]\beq
  -2\frac{{\bf r}_{12}{\bf r}_{23}}{r_{12}^2r_{23}^2}-
  2\pi (c-\ln r_{13})(\delta^2(r_{12})+\delta^2(r_{23})) -
  4\pi^2\delta^2(r_{12})\delta^2(r_{23})(\nabla_1+\nabla_3)^2\nabla_1^{-2}
  \nabla_3^{-2}]\eeq
 Here $r_{12}=r_1-r_2$ etc., $c=\ln (2/m)+\psi(1)$ where $m$ is the gluon mass
 acting as an infrared cutoff. The first two terms in $A_1$ correspond to the
 first two terms in (13). The last term in (23) corresponds to the last term
 in (13).

 The transformation of the part $G_2$ to the coordinate space encounters
 a certain difficulty in transforming the gluon Regge trajectory to the
 coordinate space, which requires introduction of an ultraviolet cutoff
 $\epsilon$. Of course the final results do not depend on $\epsilon$.
 One obtains
 \beq
 G_2(r_1,r_2,r_3)=A_2D_2(r_1,r_3)
 \eeq
 where $A_2$ is another operator in the coordinate space
 \beq
 A_2=-\frac{g^2N_c}{8\pi^3}(\frac{1}{r_{23}^2}-2\pi c\delta^2(r_{23}))
 +\delta^2(r_{23})\omega(-i\nabla_3)
     -\frac{g^2N_c}{8\pi^3}(\frac{1}{r_{12}^2}-2\pi c\delta^2(r_{12}))
     +\delta^2(r_{12})\omega(-i\nabla_1)
     \eeq
     The 4 terms in $A_2$ correspond to the 4 respective terms in $G_2$.
     The singular operators $1/r_{12}^2$ and $1/r_{23}^2$ are in fact defined
     with the help of $\epsilon$ as
     \beq
     \frac{1}{r^2}\equiv\frac{1}{r^2+\epsilon^2}+2\pi\delta^2(r)\ln\epsilon,
     \ \ \epsilon\rightarrow 0
     \eeq
     They do not depend on $\epsilon$.

     Summing $A_1$ and $A_2$ we find that the terms containing $\ln m$ cancel.
     As a result, $G(r_1,r_2,r_3)$ does not depend on the gluon mass and is
     infrared stable (which was to be expected, of course).

Now we can proceed to study the conformal invariance of the integral
\beq
I=\int d^2r_1d^2r_2d^2r_3 \Phi(r_1,r_2,r_2) G(r_1,r_2,r_3)
\eeq
where function $\Phi$ is conformal invariant.  We shall demonstrate that
the integral $I$ does not change under conformal transformations. In doing so
we shall use the fact that the BFKL solution $\Psi(r_1,r_2)=
\nabla_1^{-2}\nabla_2^{-2}D_2(r_1,r_2)$ is conformal invariant.

We shall study the behaviour of the function $G$ only under the
inversion, the invariance under translations and rescaling being obvious.
In the complex notation, under inversion 
\beq
r\rightarrow 1/r,\ \ k\equiv -i\partial\rightarrow r^2k
\eeq
from which we also conclude (for real $r$) 
\beq
d^2r\rightarrow d^2r/r^4
\eeq
and
\beq
D_2(r_1,r_2)=
r_1^4r_2^4D_2(r_1,r_2)
\eeq

Certain parts of $G$ give contributions which are evidently invariant
under inversion. Take the last term from $A_1$. It leads to an integral
\beq
I_1=\frac{g^2N}{2\pi}\int d^2r\Phi(r,r,r)\nabla^2\Psi(r,r)
\eeq
It is conformal invariant, since both functions $\Phi$ and
$\Psi$ are invariant, and the factor $r^{-4}$ from $d^2r$ is cancelled
by the factor $r^4$ from $\nabla^2$.

Terms with the denominators $r_{12}^2$ and/or $r_{23}^2$ from $A_1+A_2$
combine ito an integral
\beq
I_2=-\frac{g^2N}{8\pi^3}
\int d^2r_1d^2r_2d^2r_3\Phi(r_1,r_2,r_3)\frac{r_{13}^2}{r_{12}^2
r_{23}^2}
D_2(r_1,r_3)
\eeq
in which the regularization (26) is implied. Under inversion the
ultraviolet cutoff $\epsilon$ is transformed into
$\epsilon_1=r_1r_2\epsilon$ and $\epsilon_2=r_2r_3\epsilon$ in the
denominators $r_{12}^2$ and $r_{23}^2$ respectively. This gives rise
to a change of $I_2$ under inversion:
\beq
\Delta I_2=\frac{g^2N}{4\pi^2}\int d^2r_1d^2r_3
(\Phi(r_1,r_1,r_3)\ln r_1^2+\Phi(r_1,r_3,r_3)\ln r_3^2)D_2(r_1,r_3)
\eeq

The rest of the terms in $A_1+A_2$, proportional either to $\delta^2(r_{12})$
or to $\delta^2(r_{23})$ can be divided into two parts. The first contains
terms in which the $\delta$-function is multiplied either by a constant or by
$\ln r_{13}$ It gives rise to a part of the integral $I_3$. Terms with a
constant
are evidently invariant under inversion. However those containing
$\ln r_{13}$ are not and the corresponding change
of $I_3$  is trivially found to be
\beq
\Delta I_3=
-\frac{g^2N}{8\pi^2}\int d^2r_1d^2r_3(\Phi(r_1,r_3,r_3)+
\Phi(r_1,r_1,r_3))\ln (r_1^2r_3^2)
D_2(r_1,r_3)
\eeq

The second part contains differential operators acting on $D_2$. It has a form
\beq
-\frac{g^2N_c}{8\pi^2}(a_1\delta^2(r_{12})+a_3\delta^2(r_{23}))
\eeq
where
\beq
a_1=\nabla_1^2\ln r_{13}^2\,\nabla_1^{-2}+\ln(-\nabla_1^2)\equiv
\nabla_1^2\,\tilde{a}_1\,\nabla_1^{-2}
\eeq
and $a_3$ is obtained by interchange $1\leftrightarrow 3$. The operator $a_1$
can be transformed into a different form in which its properties under
inversion become apparent. Indeed we have, in the complex notation
\beq
\tilde{a}_1=\ln r_{13} +\ln k_1 +c.c
\eeq
One can prove [6] that
\beq
 \ln r_{13} +\ln k_1 =\ln (r_{13}^2 k_1)-k_1^{-1}\ln r_{13}\, k_1
 \eeq
from which one finds
\beq
a_1=k_1\ln (r_{13}^2k_1)\,k_1^{-1}-\ln r_{13}+c.c
\eeq
Under the inversion
\beq
\ln r_{13}^2k_1\rightarrow \ln r_{13}^2k_1-2\ln r_3
\eeq
so that the change in $a_1$ is
\beq
\Delta a_1=-2\ln r_{13}+\ln(r_1r_3)+c.c.=\ln\frac{r_1^2}{r_3^2}
\eeq
It follows that the change of the last part of the integral $I_4$,
which comes from (35), is
\beq
\Delta I_4=-\frac{g^2N}{8\pi^2}\int d^2r_1d^2r_3
(\Phi(r_1,r_1,r_3)-\Phi(r_1,r_3,r_3))D_2(r_1,r_3)\ln\frac{r_1^2}{r_3^2}
\eeq

In the sum all the changes cancel: $\Delta I_2 +\Delta I_3 +\Delta I_4 =0$,
which proves that the total integral $I$ is indeed invariant under inversion.

\section{Double pomeron exchange vs. triple pomeron}
Let us return to the 4-gluon equation for the diffractive amplitude
\beq
S_{40}D_4^{(0)}=D_{40}^{(0)}+ZD_2+g^2N_c(V_{12}+V_{34})D_4^{(0)}
\eeq
The solution of this equation is a known function, constructed as a sum
of the DPE and a TPI parts as in Eq. (14).
Each part consists of several terms, corresponding to various terms
in $D_{40}^{(0)}$ and $ZD_2$ (see Eqs. (8) and (10)).

Let us separate from this known exact solution some arbitrary function
$f$, which may depend on the angular momentum $j$:
\beq
D_4^{(0)}(j)=f(j)+\tilde{D}_4^{(0)}(j)
\eeq
Putting this into (43) we obtain an equation for the new 4-gluon function
$\tilde{D}_4^{(0)}$
\beq
S_{40}\tilde{D}_4^{(0)}=D_{40}^{(0)}+ZD_2-(S_{40}-g^2N_c(V_{12}+V_{34}))f
+g^2N_c(V_{12}+V_{34})\tilde{D}_4^{(0)}
\eeq
So the inhomogeneous term has changed
\beq
D_{40}^{(0)}+ZD_2\rightarrow D_{40}^{(0)}+ZD_2-(S_{40}-g^2N_c(V_{12}+V_{34}))f
\eeq

The significance of this seemingly trivial procedure is that if one chooses
$f$ to be the BFKL function, depending on some gluon momenta, the added
term aquires a structure of the triple pomeron contribution. So the net
effect of this procedure will be to add some new triple pomeron term 
and simultaneously to add a new simpler term $f$ to the amplitude itself.
This means that one can calculate some specific triple pomeron contributions
expressing them in terms of simple functions.

Let us see how this procedure works in some important cases.
We shall consider only the forward case, for simplicity.
Let $f=(1/2)g^2D_2(1)=(1/2)g^2D_2(2+3+4)$.
We are going to calculate then
\beq
X=(1/2)g^2(S_{40}-g^2N_c(V_{12}+V_{34}))D_2(1)
\eeq
 Using Eq. (1)
we can express the $j-1$ term in $S_{40}$ in terms of
 the forward BFKL interaction.
 $V_0$ and $\omega$ to
 obtain
\beq
X=(1/2)g^2D_{20}+(1/2)g^2(g^2N_c(V_0-V_{12}-V_{34})+2\omega(1)-\sum\omega(i))
D_2(1)
\eeq
It is straightforward to find that
\beq
(V_0-V_{12})D_2(1)=-W(1,2,3+4)
\eeq
where $W$ has been defined by (12).
The bootstrap relation gives
\beq
g^2N_cV_{34}D_2(1)=2(\omega(3+4)-\omega(3)-\omega(4))D_2(1)
\eeq
Passing to functions $G(1,3)$ with two arguments defined as $G(1,2,3)$,
Eq. (11), for $1+2+3=0$ 
we finally obtain
\beq
X=(1/2)g^2(D_{20}(1)+G(1,3+4)+D_2(1)(\omega(3)+\omega(4)-2\omega(3+4))+
D_2(1+2)(\omega(2)-\omega(1+2))
\eeq

If we put this into equation (43), we find that the changed function
$\tilde{D}_4^{(0)}$ will satisfy it with a new inhomogeneous term
\[
D_{40}+ZD_2+(1/2)g^2(-D_{20}(1)-G(1,3+4)\]\beq
-D_2(1)(\omega(3)+\omega(4)-2\omega(3+4))-
D_2(1+2)(\omega(2)-\omega(1+2))
\eeq
Note that the additional term $-(1/2)g^2D_{20}$ will cancel the identical
term in $D_{40}^{(0)}$. As a result, we have converted the double pomeron 
exchange contribution coming from $(1/2)g^2D_{20}(1)$ into a triple pomeron 
contribution corresponding essentially to $G(1,3+4)$ 
plus the explicitly separated term $(1/2)g^2D_2(1)$. In other words,
one can calculate the triple pomeron contribution corresponding to a
vertex
\[
(1/2)g^2(-G(1,3+4)-D_2(1)(\omega(3)+\omega(4)-2\omega(3+4))-
D_2(1+2)(\omega(2)-\omega(1+2))
\]
as a sum of the double pomeron exchange coupled to $-(1/2)g^2D_{20}(1)$ and
a term $(1/2)g^2D_2(1)$.

Evidently this result is trivially generalized for $f=D_2(i)$, $i=2,3,4$
by simple permutation of indeces 1,2,3 and 4.

Now let us consider a case when $f=(1/2)g^2D_2(1+2)$. In this case
\beq
X=(1/2)g^2D_{20}+(1/2)g^2(g^2N_c(V_0-V_{12}-V_{34})+2\omega(1+2)-\sum\omega(i))
D_2(1+2)
\eeq
In terms of $W$ we have
\beq
V_0D_2(1+2)=-W(1+2,0,3+4)
\eeq
The bootstrap gives
\beq
g^2N_c(V_{12}+V_{34})D_2(1+2)=2D_2(1+2)(2\omega(1+2)-\sum\omega(i))
\eeq
so that in terms of $G$ we obtain
\beq
X=(1/2)g^2(-D_{20}(1+2)+G(1+2,3+4)-D(1+2)(4\omega(1+2)-\sum\omega(i)))
\eeq
Again we see that the term with the double pomeron exchange
coupled to $g^2(1/2)D_{20}(1+2)$ can be transformed into a triple pomeron
vertex, essentially, into $-G(1+2,3+4)$ term.

Finally we study a more complicated case with $f=(1/2)g^2D_2(1+3)$.
In this case we find
\beq
V_0D_2(1+3)=-W(1+3,0,2+4) 
\eeq
Calculation of $V_{12}$ or $V_{34}$ applied to $D_2(1+3)$ is done using the
formula derived in the appendix to [3]. It gives
\beq
V_{12}D_2(1+3)=W(2+4,1,3)+W(4,2,1+3)-W(3,1+2,4)-W(1+3,0,2+4)
\eeq
and
\beq
V_{34}D_2(1+3)=W(2+4,3,1)+W(2,4,1+3)-W(1,3+4,2)-W(1+3,0,2+4)
\eeq
Using these results we obtain for this case
\[
X=(1/2)g^2(D_{20}(1+3)+G(1,2+4)+G(2,1+3)+G(3,2+4)+G(4,1+3))\]\[
-G(1,2)-G(3,4)-G(1+3,2+4)-D_2(1)(\omega(3+4)-\omega(3))\]\beq
-
D_2(2)(\omega(3+4)-\omega(4))-D_2(3)(\omega(1+2)-\omega(1))-
D_2(4)(\omega(1+2)-\omega(2)))
\eeq
The result for $f=(1/2)g^2D_2(1+4)$ is obtained from this after the
permutation of 3 and 4.

Inspecting these results and comparing them with the form of our
triple pomeron vertex, we see that only four terms
\beq
(1/2)g^2(G(1,3)+G(1,4)+G(2,3)+G(2,4))
\eeq
are not changed under these transformations and thus correspond to a true
triple pomeron interaction. All the rest can be transformed into terms 
which are essentially double 
pomeron exchange contribution. Conversely, one can eliminate terms from
the double
pomeron exchange  substituting them by  equivalent triple
pomeron contributions.

The most radical result follows if one takes
\beq
f=D_{40}^{(0)}(D_{20}\rightarrow D_2)
\eeq
In this case all the double exchange becomes cancelled and the whole amplitude
is given by a sum of two terms (in an evident symbolic notation)
\beq
D_4=D_{40}(D_{20}\rightarrow D_2)+\int_0^YG_2(Y-y)G_2(Y-y)\tilde{Z}D_2(y)
\eeq
where the new vertex is found to be
\[
\tilde{Z}D_2=(1/2)g^2(G(1,3)+G(1,4)+G(2,3)+G(2,4)+G(1+2,3+4)\]\beq-
G(1,3+4)-G(2,3+4)-G(3,1+2)-G(4,1+2))
\eeq
Comparing to (20) for the forward case, we observe that
it coincides with the part $V(1234)$ of the vertex introduced in [1].

\section{Coupling to pomerons. Comparison with the dipole approach}
Using the possibility to transfer the DPE part
into the TPI one, we shall study the triple pomeron vertex
in the simpler form (20) for a non-forward direction.
In the coordinate space of 4 gluons, the dependence on only the sum of
the momenta of two gluons, say, 1+2, is translated into a factor
$\delta^2(r_{12})$, so that the two gluons have to be taken at the
same point. However the wave functions $\Psi(r_1,r_2)$
and $\Psi
(r_3,r_4)$ of the two final pomerons coupled to the vertex 
vanish if $r_1=r_2$ and $r_3=r_4$ respectively. Therefore all terms in
 Eq.(20) which depend either only on the sum 1+2 or/and only on the sum
3+4 give zero, coupled to the two pomerons.
This leaves only the four terms, corresponding to the mentioned ``true''
triple pomeron vertex
\beq
\tilde{Z}D_2=(1/2)g^2
(G(1,2+4,3)+G(1,2+3,4)+G(2,1+4,3)+G(2,1+3,4))
\eeq 
Both pomeron functions $\Psi(r_1,r_2)$ and $\Psi(r_3,r_4)$ are symmetric
in their respective arguments, due to the positive signature of the pomeron.
Therefore all terms in (65) give identical contributions so that we can take 
\beq
\tilde{Z}D_2=2g^2
G(1,2+3,4)
\eeq
Turning to the explicit expression  of $G(1,2,3)$ in the coordinate space,
found in Sec. 3, we can split it into a "proper part"
\beq
G^{pr}(r_1,r_2,r_3)=-\frac{g^2N_c}{8\pi^3}\frac{r_{13}^2}{r_{12}^2
r_{23}^2}D_2(r_1,r_3)
\eeq
(with the regularization (26) implied) and an "improper part"
including all the rest terms, proportional to $\delta^2(r_{12})$
or/and $\delta^2(r_{23})$. Noting that in the coordinate space
(66) is proportional to $\delta^2(r_{23})$, we find that in the
improper part at least three gluons, either 123 or 234, are to be
taken at the same point in the transverse space. Then these terms will
vanish due to the mentioned property of the pomeron wave function.
Therefore the final triple pomeron vertex is given by only the proper part
of $G$, Eq. (67).

Coupling this triple pomeron vertex to the two final pomerons,
we arrive at the following expression for the
triple pomeron contribution to the diffractive (non-forward) amplitude:
\beq
D_4^{TPI}(Y)=-\frac{g^4N}{4\pi^3}
\int_0^Y dy\int d^2r_1d^2r_2d^2r_3\frac{r_{13}^2}{r_{12}^2r_{23}^2}
D_2(r_1,r_3;y)\Psi_1(r_1,r_2;Y-y)\Psi_2(r_2,r_3;Y-y)
\eeq
Expressing the initial pomeron via the non-amputated function $\Psi$ we can
rewrite (68) as
\beq
D_4^{TPI}(Y)=-\frac{g^4N}{4\pi^3}\int_0^Y dy
\int \frac{d^2r_1d^2r_2d^2r_3}{r_{13}^2r_{12}^2r_{23}^2}
\Psi_1(r_1,r_2;Y-y)\Psi_2(r_2,r_3;Y-y)r_{13}^4\nabla_1^2\nabla_3^2
\Psi(r_1,r_3;y)
\eeq
In this form it is evident that the triple pomeron vertex is not
symmetric with respect to the initial pomeron  and two final ones:
there appears an extra operator $r_{13}^4\nabla_1^2\nabla_3^2$ acting on
the initial pomeron. Note that this operator is essentially a product of
the Kasimir operators of the conformal group for the holomorphic and
antiholomorphic parts:
\beq
M^2(1,3)\bar{M}^2(1,3)=(1/16)r^4_{13}\nabla_1^2\nabla^2_3
\eeq
So one expects simplifications to occur provided one passes to the conformal
basis for the BFKL solutions.

This basis is formed by functions (in complex notation)
\beq
E_{n,\nu,r_0}(r_1,r_2)\equiv E_{\mu}(r_1,r_2)
=(\frac{r_{12}}{r_{10}r_{20}})^{\frac{1-n}{2}+i\nu}
(\frac{r^*_{12}}{r^*_{10}r^*_{20}})^{\frac{1+n}{2}+i\nu}
\eeq
They are proper functions of (70)
In fact,
\beq
M^2\bar{M}^2\,E_{\mu}(r_1,r_2)=\frac{\pi^8}{4}\frac{E_{\mu}(r_1,r_2)}
{a_{n+1,\nu}a_{n-1,\nu}}
\eeq
where we use the standard notation
\beq
a_{n,\nu}\equiv a_{\mu}=\frac{\pi^4}{2}\frac{1}{\nu^2+n^2/4}
\eeq
They form a complete system:
\beq
r_{12}^4\delta^2(r_{11'})\delta^2(r_{22'})=
\sum_{\mu}
E_{\mu}(r_1,r_2)E^*_{\mu}(r'_1,r'_2)
\eeq
where we use a notation
\beq
\sum_{\mu}=\sum_{n=-\infty}^{\infty}\int d\nu\frac{1}{2a_{n,\nu}}
 \int d^2r_0
\eeq
and satisfy the orthogonality relation
\[
\int\frac{d^2r_1d^2r_2}{r_{12}^4}
E_{n,\nu,r_0}(r_1,r_2)E^*_{n',\nu',r'_0}(r_1,r_2)\]\beq
=a_{n,\nu}\delta_{nn'}\delta(\nu-\nu')\delta^2(r_{00'})+
b_{n\nu}\delta_{n,-n'}\delta(\nu+\nu')|r_{00'}|^{-2-4i\nu}(\frac{r_{00'}}
{r^*_{00'}})^n
\eeq
The coefficients $b_{n\nu}$ may be found in [7].
Using these properties one can express the pomeron wave function as
\beq
\Psi(r_1,r_2;y)=\sum_{\mu}e^{y\omega_{\mu}}E_{\mu}(r_1,r_2)
\langle\mu|\Psi_0\rangle
\eeq
where $\omega_{\mu}=\omega_{n\nu}$ are the eigenvalues of the BFKL kernel
and we have defined
\beq
\langle \mu|\Psi_0\rangle=
\int \frac{d^2r_1d^2r_2}{r_{12}^4}E^*_{\mu}(r_1,r_2)\Psi_0(r_1,r_2)
\eeq

We present all the three pomerons in (69) as a superposition (77) of
 conformal states. As mentioned, the operator acting on the initial pomeron
is essentially
a product of Kasimir operators, so we can use Eq. (70). We then obtain,
after the integration over $y$
\[
D_4^{TPI}(Y)=-\frac{g^4N}{4\pi^3}\sum_{\mu,\mu_1,\mu_2}
\langle\mu|\Psi_0\rangle\langle\mu_1|\Psi_{10}\rangle
\langle\mu_2|\Psi_{20}\rangle
\frac{e^{Y(\omega_{\mu_1}+\omega_{\mu_2})}-e^{Y\omega_{\mu}}}
{\omega_{\mu_1}+\omega_{\mu_2}-\omega_{\mu}}
\]\beq
\frac{4\pi^8}{a_{n-1,\nu}a_{n+1,\nu}}
\int \frac{d^2r_1d^2r_2d^2r_3}{r_{13}^2r_{12}^2r_{23}^2}
E_{\mu_1}(r_1,r_2)E_{\mu_2}(r_2,r_3)E_{\mu}(r_1,r_3)
\eeq
In this form the triple pomeron contribution can be compared to the 
double dipole density found by Peschanski [4] in  A.H.Mueller's
colour dipole approach. One  observes
that the two expressions differ only in the sign and factor 
\[
\frac{4\pi^8}{a_{n-1,\nu}a_{n+1,\nu}}
\]
which in our approach distinguishes the initial pomeron from the two
final ones. The integral over the coordinates of the three pomerons is
the same. So essentially the three-pomeron contribution to the diffractive
amplitude found in our approach coincides with the double dipole density
in the dipole approach.

However one should not forget that in our $s$-channel unitarity approach
the TPI term (79) does not exhaust all the diffractive amplitude. From the
derivation of Sec. 4 it follows
\beq D_4^{(0)}=D_{40}(D_{20}\rightarrow D_2)+D_4^{TPI}\eeq
At high energies the TPI term behaves essentially as $s^{2\Delta}$ and the first one
as $s^{\Delta}$ where $\Delta$ is the BFKL intercept. So one might think
that the first term could be neglected. However the correct region of
the validity of the leading log approximation, implicit in the hard
pomeron theory, is $g^2\ln s\sim 1$ when the two terms in (80) have the
same order of magnitude. The dipole approach uses essentially the same
leading log approximation. Therefore the fact that it leads to the
double dipole density which coincides only with the TPI term in
the $s$-channel unitarity approach and shows no trace of the first term
points to certain differences between the two approaches.

\section{Higher-order densities in the dipole approach}
In the  colour dipole formalism the k-fold inclusive dipole
density is obtained
as the $k$-th functional derivative of the functional $D\{u(r_i,r_f)\}$,
taken at
$u(r_i,r_f)=1$ [8]. The arguments $r_i$ and
$r_f$ are the dipole endpoints in the transverse plane. In the following
in many cases we
denote them as a single variable $r$ for brevity.
The functional $D$ satisfies a simple equation
\[
D(r_1,r_0,y,u)=u(r_1,r_0)e^{2y\omega(r_{10})}\]\beq+
\frac{g^2N_c}{8\pi^3}
\int_0^y dy' e^{2(y-y')\omega(r_{10})}\int d^2r_2\frac{r_{10}^2}
{r_{12}^2r_{20}^2}D(r_1,r_2,y',u)D(r_2,r_0,y',u)
\eeq
Here $r_1$ and $r_0$ are the end points of the $q\bar q$ pair which
determine the initial dipole;
$\omega(r)$ {\it is not} a Fourier
 transform of the trajectory,
but just $\omega(k)$ with $k/m$ formally substituted by $r/\epsilon$,
where $\epsilon$ is an ultraviolet cutoff. This cutoff is also implied in
the singular kernel of the integral operator in $r_2$.
Eq. (81) is compatible with the normalization condition
$
D(u=1)=1
$.

Taking the $k$-th derivative we arrive at an equation for the $k$-fold
dipole density. For $k>1$ we obtain
\[
n_k(r_1,r_0;\rho_1,...\rho_k;y)=
\frac{g^2N_c}{8\pi^3}
\int_0^y dy' e^{2(y-y')\omega(r_{10})}\int d^2r_2\frac{r_{10}^2}
{r_{12}^2r_{20}^2}n_k(r_1,r_2;\rho_1,...\rho_k;y')+(r_1 \leftrightarrow r_0)\]\[
+\frac{g^2N_c}{8\pi^3}
\int_0^y dy' e^{2(y-y')\omega(r_{10})}\int d^2r_2\frac{r_{10}^2}
{r_{12}^2r_{20}^2}\sum_{l=1}^{k-1}(n_l(r_1,r_2;\rho_1,...\rho_l;y')
n_{k-l}(r_2,r_0;\rho_{l+1},...\rho_k;y')\]\beq+\ symmetrization\ terms)
\eeq
where the symmetrization terms (ST) are obtained from the explicitly shown one
by taking all different divisions  of arguments $\rho_1,...\rho_k$ into
two groups with $l$ and $l-k$ arguments.
For $k=1$ an inhomogeneous term should be added whose form is clear from (81).
One should note that the operator on the right-hand side acts nontrivially 
only on the first argument of the density $n_k$. Its action on the rapidity
variable, on the contrary, is rather simple. Multiplying the
equation by $ e^{-2y\omega(r_{10})}$, differentiating then with respect
to $y$
and passing  to the $j$-space
by the standard Mellin transformation
 one obtains
\[
(j-1) n_k(r_1,r_0;\rho_1,...\rho_k;j)=
\frac{g^2N_c}{4\pi^3}
\int d^2r_2L(r_1,r_2,r_{20})n_k(r_1,r_2;\rho_1,...\rho_k;y)\]\[+
\frac{g^2N_c}{8\pi^3}\int\frac{dj_1dj_2}{(2\pi i)^2(j+1-j_1-j_2)}
\int d^2r_2\frac{r_{10}^2}
{r_{12}^2r_{20}^2}\]\beq
\sum_{l=1}^{k-1}(n_l(r_1,r_2;\rho_1,...\rho_l;j_1)
n_{k-l}(r_2,r_0;\rho_{l+1},...\rho_k;j_2)+\ ST)
\eeq
where we introduced the BFKL kernel in the coordinate
space
\beq
L(r_{12},r_{20})=\frac{r_{10}^2}
{(r_{12}^2+\epsilon^2)(r_{20}^2+\epsilon^2)}-2\pi\ln\frac
{r_{10}}{\epsilon}(\delta^2(r_{12})+\delta^2(r_{20})\eeq
Comparing with (26) we see that it does not depend on $\epsilon$ and
is ultraviolet stable.

To solve this equation we present
 the dependence
of the densities on  their first two arguments  in  the conformal
basis:
\beq
n_k(r_1,r_0)=\sum_{\mu}E_{\mu}(r_1,r_0)n_k^{\mu}
\eeq
Here we have suppressed all other arguments in $n_k$ irrelevant for the
time being. The densities $n_k^{\mu}$ in a given conformal state are
obtained from $n_k(r_1,r_0)$ by the inverse transformation which follows
from property (76) and a relation between $E_{n,\nu}$ and $E_{-n,-\nu}$
(see [7])
\beq
n_k^{\mu}=\int \frac{d^2r_1d^2r_0}{r_{10}^4}E^*_{\mu}(r_1,r_0)
n_k(r_1,r_0)
\eeq

So, to pass to the conformal basis, we integrate Eq. (83) over $r_1$ and
 $r_0$ as
indicated in (86). The first term on the right-hand side can be simplified
due to the property of the BFKL kernel
\beq
\frac{g^2N_c}{4\pi^3}
\int d^2r_2L(r_{12},r_{20})E_{\mu}(r_1,r_2)=\omega_{\mu}E_{\mu}(r_1,r_0)
\eeq
Therefore after the integration we obtain 
\[
(j-1-\omega_{\mu})n_k^{\mu}(\rho_1,...\rho_k,j)=
\frac{g^2N_c}{8\pi^3}
\int\frac{dj_1dj_2}{(2\pi i)^2(j+1-j_1-j_2)}\]\beq
\int \frac{d^2r_1d^2r_2d^2r_0}
{r_{12}^2r_{20}^2r_{10}^2}E^*_{\mu}(r_{10})
\sum_{l=1}^{k-1}(n_l(r_1,r_2;\rho_1,...\rho_l;j_1)
n_{k-l}(r_2,r_0;\rho_{l+1},...\rho_k;j_2)+\ ST)
\eeq
To find the final form of the equation we have only to present also
the densities $n_l$ and $n_{k-l}$ as functions of their first arguments
in the form (86). Then we get
\[
(j-1-\omega_{\mu})n_k^{\mu}(\rho_1,...\rho_k,j)=
\int\frac{dj_1dj_2}{(2\pi i)^2(j+1-j_1-j_2)}\]\beq
\sum_{\mu_1,\mu_2}V_{\mu,\mu_1\mu_2}
\sum_{l=1}^{k-1}(n_l^{\mu_1}(\rho_1,...\rho_l;j_1)
n_{k-l}^{\mu_2}(\rho_{l+1},...\rho_k;j_2)+\ ST)
\eeq
where
\beq
V_{\mu\mu_1\mu_2}=\frac{g^2N_c}{8\pi^3}\int \frac{d^2r_1d^2r_2d^2r_0}
{r_{12}^2r_{20}^2r_{10}^2}E^*_{\mu}(r_{10})E_{\mu_1}(r_{12})E_{\mu_2}(r_{20})
\eeq
is just one half of the three-pomeron vertex introduced by Peschanski.

Eq. (89) allows to obtain succesively dipole densities for any number of
dipoles starting from the lowest order one-dipole density, for which
\beq
n_1^{\mu}(\rho)=\frac{E^*_{\mu}(\rho)}{j-1-\omega_{\mu}}
\frac{1}{\rho^4}
\eeq
(we recall that in this notation $\rho$ includes two endpoints of the
dipole $\rho_i$ and $\rho_f$; $\rho^2\equiv \rho_{if}^2$).

Putting this into (89) for $k=2$ and integrating over $j_1$ and $j_2$
 we arrive at the
expression obtained by Peschanski
\beq
n_2^{\mu}(\rho_1,\rho_2;j)=\frac{1}{\omega-\omega_{\mu}}
\sum_{\mu_1,\mu_2}\frac{1}{\omega-\omega_{\mu_1}-
\omega_{\mu_2}}V_{\mu,\mu_1,\mu_2}E^*_{\mu_1}(\rho_1)E^*_{\mu_2}(\rho_2)
\frac{1}{\rho_1^4\rho_2^4}
\eeq
where $\omega=j-1$. (To compare with [5] one should take into acount that
factors $1/(2a_{\mu})$ are included in the definition of sums over $\mu$'s
in our notation).

Now we continue this process and study the density for three dipoles.
Eq. (89) for $k=3$ reads
\[
n_3^{\mu}(\rho_1,\rho_2,\rho_3;j)=\frac{1}{
\omega-\omega_{\mu}}
\int\frac{dj_1dj_2}{(2\pi i)^2(j+1-j_1-j_2)}\]\beq
\sum_{\mu_1,\mu_2}V_{\mu,\mu_1\mu_2}
(n_1^{\mu_1}(\rho_1;j_1)
n_{2}^{\mu_2}(\rho_2,\rho_3;j_2)+\ ST)
\eeq
Let us study the term written explicitly. We put in it the expressions for
$n_1^{\mu_1}$ and $n_2^{\mu_2}$ obtained earlier. Then we get,
after integrations over $j_1$ and $j_2$:
\[
n_3^{\mu}(\rho_1,\rho_2,\rho_3;j)= \frac{1}{
\omega-\omega_{\mu}}
\sum_{\mu_1,\mu_2,\mu_3,\mu_4}V_{\mu,\mu_1\mu_2}V_{\mu_2\mu_3\mu_4}\]\beq
\frac{1}{(\omega-\omega_{\mu_1}-\omega_{\mu_2})
(\omega-\omega_{\mu_1}-\omega_{\mu_3}-\omega_{\mu_4})}
E^*_{\mu_1}(\rho_1)E^*_{\mu_3}(\rho_2)E^*_{\mu_4}(\rho_3)
\frac{1}{\rho_1^4\rho_2^4\rho_3^4}
\eeq
To this term we have to add terms which symmetrize in the three dipoles.

Studying (94) we see that it corresponds to the picture when first the
initial pomeron splits into two pomerons, 1 and 2, and afterwards the pomeron 2
splits into pomerons 3 and 4. One does not find here a local  vertex
for the transition of the initial pomeron into three final ones. It is not
difficult to see under which condition one would get such a local vertex.
If we forget about the dependence of the second denominator on $\mu_2$,
then one can sum over  $\mu_2$. Using  (74)
one obtains
\beq
\sum_{\mu_2}V_{\mu\mu_1\mu_2}V_{\mu_2\mu_3\mu_4}=\left(\frac{g^2N_c}{8\pi^3}
\right)^2
\int\frac{d^2r_1d^2r_2d^2r_3d^2r_0}{r_{12}^2r_{10}^2r_{23}^2r_{30}^2}
E^*_{\mu}(r_1,r_0)E_{\mu_1}(r_1,r_2)E_{\mu_3}(r_2,r_3)E_{\mu_4}(r_3,r_0)
\eeq 
which is just the vertex from one to three pomerons introduced by Peschanski
in [5].
However, the described summation is not possible due to the second
denominator. It implies that the pomeron 2 has to evolve in $y$ from
the point of its formation from the initial pomeron up to the
point of its splitting into the final pomerons 3 and 4.

Thus our conclusion is that the vertex for transition from 1 to $k$ pomerons
introduced by Peschanski, in fact, does not appear in the solution of
the Mueller
equation for the $k$-fold density, which rather corresponds to a set of all
fan diagrams with only the triple pomeron coupling. Absence of higher-order
couplings can be directly traced to the structure of the equation (81)
for the generating functional $D$, quadratic in $D$.

To conclude this section we note that at asymptotic energies the higher-
order densities in the dipole approach correspond to the standard
Regge-Gribov picture , in the
tree approximation (fan diagrams), with only the triple pomeron interaction,
which however has a highly complicated non-local form.
Indeed, the triple pomeron interaction present in (94) corresponds to a
structure
\beq
T=\frac{g^2N_c}{8\pi^3}
\int\frac{d^2r_1d^2r_2d^2r_3}{r_{12}^2r_{23}^2r_{31}^2}
\tilde{G}_3(r_1,r_2;r_1',r'_2)\tilde{G}_1(r_2,r_3;r'_2,r'_3)
\tilde{G}_2(r_3,r_1;r'_3,r'_1)
\eeq
where $\tilde{G}_i,\ i=1,2,3$ are Green functions of the interacting
pomerons defined as
\beq
\tilde{G}(r_1,r_2;r'_1,r'_2)=
\sum_{\mu}\frac{E_{\mu}(r_1,r_2)E^*_{\mu}(r'_1,r'_2)}
{\omega-\omega_{\mu}}
\eeq
They are not the physical BFKL Green functions. The latter  include an extra
factor depending on $\mu$:
\beq
G(r_1,r_2;r'_1,r'_2)=\frac{1}{4\pi^8}
\sum_{\mu}a_{n+1,\nu}a_{n-1,\nu}\frac{E_{\mu}(r_1,r_2)E^*_{\mu}(r'_1,r'_2)}
{\omega-\omega_{\mu}}
\eeq
However
in the limit $s\rightarrow\infty$ only the lowest conformal
weights contribute $n=\nu=0$ for which $a_{n\pm 1,\nu}=2\pi^4$
and (97) and (98) coincide. Then we can forget about tildas in (96).

We transform the Green functions to  given total momenta of the pomerons
presenting
\beq
G_3(r_1,r_2;r'_1,r'_2)=\int \frac{d^2l_3}{(2\pi)^2}e^{il_3(R_3-R'_3)}
G_{l_3}(r_{12},r'_{12})
\eeq
where $R_3=(1/2)(r_1+r_2)$ and similarly for the two other Green functions.
Introducing $R=r_1+r_2+r_3$ we transform the integration over the coordinates
as follows
\beq
\int d^2r_1d^2r_2d^2r_3=
\int d^2Rd^2r_{12}d^2r_{23}d^2r_{31}
\delta^2(r_{12}+r_{23}+r_{31})
\eeq
The coordinates themselves are 
$r_1=(1/3)(R-r_{21}-r_{31})$ etc., wherefrom we find
\beq
R_1=(1/6)(2R-r_{12}-r_{13}),\ R_2=(1/6)(2R-r_{21}-r_{23}),\ 
R_3=(1/6)(2R-r_{31}-r_{32})
\eeq
and
\beq
 i\sum_{j=1}^3l_jR_j=i(1/3)R\sum_{j=1}^3l_j-
i(1/6)(r_{12}l_{12}+r_{23}l_{23}+r_{31}l_{31})
\eeq
where we denoted $l_{12}=l_1-l_2$ etc.
The integral over $R$ gives $9(2\pi)^2\delta^2(l_1+l_2+l_3)$.
 Presenting  the remaining
$\delta$-function in (100) as an integral over an auxiliary momentum $q$ we
find an expression (for fixed $l_1,l_2$ and $l_3$)
\[\frac{g^2N_c}{8\pi^3}
\int\frac{d^2q}{(2\pi)^2}\frac{d^2r_{12}d^2r_{23}d^2r_{31}}
{r^2_{12}r^2_{23}r^2_{31}}\exp \left(ir_{12}(q-\frac{1}{6}l_{12})+
ir_{23}(q-\frac{1}{6}l_{23})+ir_{31}(q-\frac{1}{6}l_{31}\right)\]\beq
G_{l_3}(r_{12},r'_{12})G_{l_1}(r_{23},r'_{23})G_{l_2}(r_{31},r'_{31})
\eeq

At this point we recall the expression for the BFKL Green function with
a fixed total momentum:
\beq
G_l(r,r')=\frac{1}{(2\pi)^4}\int \frac{\nu^2 d\nu}{(\nu^2+1/4)^2}s^{\omega(\nu)}
E^{(l)}_\nu(r)E^{(l)}_\nu(r')
\eeq
where
\beq
E^{(l)}_\nu(r)=\int d^2R\exp (ilR)
\left(\frac{r}{|R+r/2||R-r/2|}\right)^{1+2i\nu}
\eeq
and where we retained only the dominant isotropic term. At $s\rightarrow\infty$
the vicinity of $\nu=0$ gives the dominant contribution. If $l\neq 0$ then the
integral in (105) converges at large $R$ and we can take the functions $E$
 out of the integral over $\nu$ at $\nu=0$. Taking then the asymptotics of the
remaining integral, we find
\beq
G_l(r,r')\simeq\frac{1}{2\pi^4}s^{\Delta}\frac{\sqrt{\pi}}{(a\ln s)^{3/2}}
E^{(l)}_0(r)E^{(l)}_0(r')
\eeq
where $\Delta=\omega_{n=0,\nu=0}$ is the BFKL intercept and
$a=7g^2N_c\zeta(3)/(2\pi^2)$.
As we see, the Green function asymptotically factorizes in $r$ and $r'$.
This means that we obtain a quantum field theory of pomerons with a 
propagator
\beq
P(y,l)=\frac{2}{\pi^2}e^{y\Delta}\frac{\sqrt{\pi}}{(ay)^{3/2}}
\eeq
(not really depending on the momentum $l$) and an interaction vertex
\[
V(l_1,l_2,l_3)= \frac{9g^2N_c}{8\pi^3}
\int\frac{d^2q}{(2\pi)^8}\frac{d^2r_{12}d^2r_{23}d^2r_{31}}
{r^2_{12}r^2_{23}r^2_{31}}\]\beq
\exp \left(ir_{12}(q-\frac{1}{6}l_{12})+
ir_{23}(q-\frac{1}{6}l_{23})+ir_{31}(q-\frac{1}{6}l_{31}\right)
E^{(l_3)}_0(r_{12})E^{(l_1)}_0(r_{23})E^{(l_2)}_0(r_{31})
\eeq

The vertex  factorizes under the sign of the integration over $q$:
\beq
V(l_1,l_2,l_3)=\frac{9g^2N_c}{8\pi^3}
\int\frac{d^2q}{(2\pi)^2}
J(l_3,q-\frac{1}{6}l_{12})J(l_1,q-\frac{1}{6}l_{23})
J(l_2,q-\frac{1}{6}l_{31})
\eeq
where
\beq
J(l,q)=\int\frac{d^2r}{(2\pi)^2 r^2}e^{iqr}E^{(l)}_0(r)
=\int\frac{d^2p}{2\pi p}\frac{1}{|p-q+l/2||p-q-l/2|}
\eeq

Note that for $l=0$ this derivation is incorrect. Calculations show
that in this case Eq. (109) for the vertex remains valid with
\beq
J(0,q)=\frac{1}{9q}
\eeq
However the Green function (104) at $l=0$ has an asymptotics
\beq G_0(r,r')\simeq\frac{1}{2\pi^2}s^{\Delta}\sqrt{\frac{\pi}{a\ln s}}
rr'\exp\left(-\frac{\ln^2(r/r')}{a\ln s}\right)
\eeq
so that the factorization is lost.

\section{Conclusions}
Study of the 4-gluon system at $N_c\rightarrow\infty$ shows that in the
leading order the system reduces to a single pomeron, as pointed out in
[4]. The diffractive amplitude, subleading in $1/N_c$, turns out to be a sum
of the DPE and TPI contributions. The triple pomeron vertex which appears
in the latter is different from the one introduced in [1] for $N_c=3$.
However it is also conformal invariant. Moreover functions $G$ out of which
both verteces are constructed are conformal invariant by themselves.

A novel feature of the diffractive amplitude in  the
hard pomeron model is the equivalence of the DPE and certain terms of TPI.
This allows to eliminate the DPE contribution altogether and substitute it by
additional TPI terms. The price of such a substitution is the appearance
of some extra terms of a structure different from both the DPE and TPI,
which are absent in the old Regge-Gribov theory.

The triple pomeron vertex obtained after this substitution coincides with
a part $V(1234)$ of the vertex introduced in [1], leading in the high-
colour limit. Coupling this vertex to two pomerons, most of the term
vanish and the rest simple expression coincides with the triple pomeron
vertex found by Peschanski in [5]. So there is a complete agreement
between the results of [1,3,5] in this respect.
However the mentioned extra terms do not
seem to appear in the dipole picture, which points to a certain difference
between this approach and the $s$-channel unitarity one.

In the dipole approach the higher-order dipole densities are found to
be represented by  a set of pomeron fan diagrams with only a triple
pomeron coupling. Four and more pomeron coupling do not appear, which
is a clear prediction of the dipole picture. It would be interesting to
verify this prediction in the framework of the $s$-channel unitarity
approach by studying transitions from 1 to 3 pomerons.

\section {Acknowledgments.}

The authors express their deep gratitude to Prof G.Venturi for his
constant interest in this work and helpful discussions.
M.A.B. thanks the INFN and University of Bologna for their hospitality
and financial help during his stay
in Bologna. 

\section{Appendix: Quark loop for non-zero momentum transfer}

Consider a $q\bar q$ loop for the scattering of a virtual photon
$\gamma^*(q)+...\rightarrow\gamma^*(q+l)+...$, $q^2=-Q^2$. The momentum
transfer $l$ is taken to be pure transversal. Then a straightforward
calculation gives for the function $f(1,2)$ corresponding to the loop
with gluon 1 attached to $q$ and gluon 2 attached to $\bar q$ the following
expression
\beq
f(k_1,k_2)=e_f^2\int_0^1d\alpha\int\frac{d^2k}{(2\pi)^3}\frac{N}{D}
\eeq
where
\beq
D=(\epsilon^2+k^2)(\epsilon_1^2+(k+k_1-\alpha l)^2)
\eeq
\beq
\epsilon^2=Q^2\alpha(1-\alpha)+m_f^2,\ \
\epsilon^2_1=(Q^2+l^2)\alpha(1-\alpha)+m_f^2
\eeq
$e_f$ and $m_f$ are the quark electric charge and mass and the numerator
for a transversal photon is
\beq
N^{T}=m_f^2 +(\alpha^2+(1-\alpha)^2){\bf k}({\bf k+k_1})-\alpha^2{\bf kl}
\eeq
and for a longitudinal photon is
\beq
N^{L}=4Q^2\alpha^2(1-\alpha)^2
\eeq

This expression can be conveniently represented as an integral over the
colour dipole density $\rho$ created by the $q\bar q$ pair at a given
distance in the transverse space:
\beq
f(k_1,k_2)=\int d^2r\rho_l(r)e^{ik_1r}
\eeq
>From (113) - (116) one finds for the transverse and longitudinal photons:
\[
\rho^T_l(r)=\frac{e_f^2}{(2\pi)^3}e^{-i\alpha lr}\int_0^1d\alpha
[m_f^2K_0(\epsilon r)K_0(\epsilon_1 r)+\]\beq
(\alpha^2+(1-\alpha)^2)\epsilon\epsilon_1 K_1(\epsilon r)K_1(\epsilon_1 r)-
\alpha(1-\alpha)(1-2\alpha)\frac{i\epsilon{\bf lr}}{r}K_0(\epsilon_1 r)
K_1(\epsilon r)]
\eeq
\beq
\rho^L_l(r)=\frac{4e_f^2Q^2}{(2\pi)^3}e^{-i\alpha lr}\int_0^1d\alpha
\alpha^2(1-\alpha)^2K_0(\epsilon r)K_0(\epsilon_1 r)
\eeq
\newpage

\section{References}

1. J.Bartels and M.Wuesthoff, Z.Phys. {\bf C 66} (1995) 157.\\
2. J.Bartels, L.N.Lipatov and M.Wuesthoff, Nucl. Phys. {\bf B464} (1996)
298.\\
3. M.A.Braun, S.Petersburg preprint SPbU-97-9 (hep-ph/9706373).\\
4. M.A.Braun, Z.Phys. {\bf C71} (1996)601.\\
5. R.Peschanski, Saclay preprint, (hep-ph/9704342); Phys.Lett.
 {\bf 409} (1997) 491.\\
6. M.A.Braun, Santiago University preprint US-FT/4-94.\\
7. L.N.Lipatov, in "Perturbative QCD", ed. A.H.Mueller (World Sci.,
Singapore, 1989).\\
8. A.H.Mueller, Nucl.Phys. {\bf B415} (1994) 373; {\bf B437} (1995) 107;\\
A.H.Mueller and B.Patel, Nucl. Phys. {\bf B425} (1994) 471.\\

\end{document}